# Reflection suppression by the hyperbolic-medium antennas and silicon particles

V.E. Babicheva
College of Optical Sciences, University of Arizona, Tucson, AZ, USA
vbab.dtu@gmail.com

**Abstract.** Optical antennas made out of materials with hyperbolic dispersion is an alternative approach to realizing efficient subwavelength scatterers and may overcome limitations imposed by plasmonic and all-dielectric designs. Recently emerged natural hyperbolic material hexagonal boron nitride supports phonon-polariton excitations with low optical losses and high anisotropy. Here we study scattering properties of the hyperbolic-medium (HM) antennas, and in particular, we consider a combination of two types of the particles - HM bars and silicon spheres - arranged in a periodic array. We analyze excitation of electric and magnetic resonances in the particles and effect of their overlap in the array. We theoretically demonstrate that decrease of reflectance from the array can be achieved with appropriate particle dimensions where electric and magnetic resonances of different particle types overlap, and the resonance oscillations are in phase. In this case, generalized Kerker condition is satisfied, and particle dimers in the array efficiently scatter light in the forward direction. The effect can be used in designing metasurfaces based on hexagonal boron nitride scatterers with an application in mid-infrared photonics.

## Introduction

Subwavelength plasmonic structures have been extensively studied for a wide range of functionalities aiming to be applied as optical antennas and enhance light-matter interaction, light harvesting, directional scattering, high-resolution near-field microscopy, to name but a few [1-8]. Later, high-index dielectric antennas have been demonstrated as a promising alternative to plasmonic particles [9-13]. Recently, we have proposed using antennas made out of natural hyperbolic material (hexagonal boron nitride), and have shown this medium is a favorable counterpart to plasmonic and all-dielectric materials for realizing optical antennas and scattering elements in metasurfaces [14]. In the hyperbolic medium, permittivity tensor components have the different signs of a real part causing hyperbola-like wave dispersion [15,16]. Medium with hyperbolic dispersion (hyperbolic medium, HM) supports propagating waves with high wavenumber, which results in such phenomena as strong and broadband spontaneous emission enhancement, anomalous heat transfer, etc. [17-22]. Furthermore, highly subwavelength confinement of optical mode in the hyperbolic medium has been demonstrated in waveguides [23-26] and tapers [27]. For arrays of particles, lattice resonances in periodical arrangement have been studied for both plasmonic and all-dielectric particles [28-33]. In our previous work, we have shown that in a similar way, particle out of hyperbolic medium can possess resonances at the wavelength defined by the period of the structure [14]. There, hexagonal boron nitride particles have been proposed as subwavelength resonators, and the possibility of multipole resonance excitations have been pointed out with an emphasis on scattering properties of particles. Most importantly, resonant overlaps of the particle multipoles cause directional scattering (Kerker effect [34-38]), and periodic arrangement of the particles results in pronounced lattice resonances.

## Results

In this work, we consider an array of particles which include both HM bars and silicon spheres. For the material with hyperbolic dispersion, we choose $\varepsilon_x = \varepsilon_y = -14.6+1i$ and $\varepsilon_z = 2.7$, which corresponds to the hexagonal boron nitride at wavelength 7 μm according to permittivity model of [39]. For the HM bar with dimensions $a_x = 0.9$ μm, $a_y = 0.5$ μm, and $a_z = 2$ μm (fixed throughout the paper), electric multipole resonance (HM-EMR) is at $\lambda_{EMR} \approx 6.2$ μm and magnetic and electric quadrupole resonances (HM-MR/EQR) is at $\lambda_{MR/EQR} \approx 4.8$ μm. In the HM antennas, higher multipole resonances are excited at the larger wavelength with respect to dipole resonances, and for detailed analysis of the modes in HM antennas, we refer to our earlier work [14]. In contrast, for a silicon sphere, higher multipole resonances are excited at the blue side from electric and magnetic dipole resonances. Upon increase of sphere radius, the resonances experience redshifts, and for the sphere radius $R = 0.5 - 1.1$ μm, they are in the wavelength range $\lambda \approx 4 - 8$ μm.

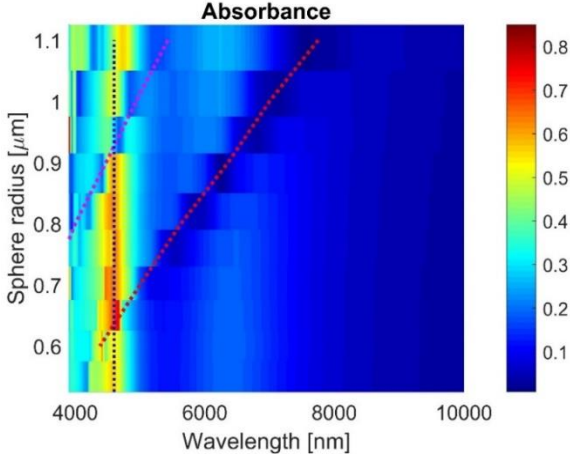

Figure 1. Absorption in HM in the array of HM bars and silicon spheres of varied radius $R = 0.5 - 1.1$ μm. Dotted red and magenta lines mark magnetic Si-MD and electric Si-ED dipole resonances of silicon sphere, respectively. The dotted blue line corresponds to magnetic and electric quadrupole resonances (HM-MR/EQR) of the HM bar. HM bar dimensions are $a_x = 0.9$ μm, $a_y = 0.5$ μm, and $a_z = 2$ μm.

Optical properties of an array that combines HM bars and silicon spheres have spectral features of both types of the particles. Variations in the sphere radius cause shifts of magnetic and electric dipole resonances and their overlap with EMR and MR/EQR of HM bar; and the sphere resonances appear as decreases in absorption in HM (Figure 1). One can map particle resonances pronounced in absorption on reflectance spectra. In Figure 2, we observe that increase of sphere radius causes a redshift of both peaks and dips in reflectance spectra, but they do not exactly match sphere's electric and magnetic dipole resonance positions. One can see that reflectance peak appears on the blue side of Si-MD and dip on the red side of Si-MD.

Furthermore, a broad reflectance dip is observed between resonances for $R = 0.75 - 0.95$ μm, when both Si-MD and Si-ED are in the proximity of HM-MR/EQR.

To clarify an origin of the reflection suppression, one can consider an example with sphere radius $R = 0.82$ μm shown in Figure 3. For $\lambda \gtrsim 6$ μm (the red side of Si-MD), both Si-MD and HM-EMR are in phase, and near-zero reflection from the array is caused by the predominantly forward scattering from the particle dimers. A similar effect is observed in the wavelength of HM-MR/EQR where the proximity of both Si-ED and Si-MD results in a band of resonant reflection suppression. In both cases, a near-zero reflectance means that a generalized Kerker condition is satisfied and dimers effectively serve as Huygens' elements.

## Conclusion

To sum up, we have studied an array of particles that includes both HM bars and silicon spheres. We have shown that overlap of electric and magnetic resonances of the particles results in a near-zero reflectance band with resonant scattering forward by the particle dimer. It reproduces Kerker effect with a generalized condition satisfied by different multipoles. This effect facilitates further development of functional metasurfaces and improvements of optical devices in terms of reflection decrease, miniaturization, and higher energy efficiency.

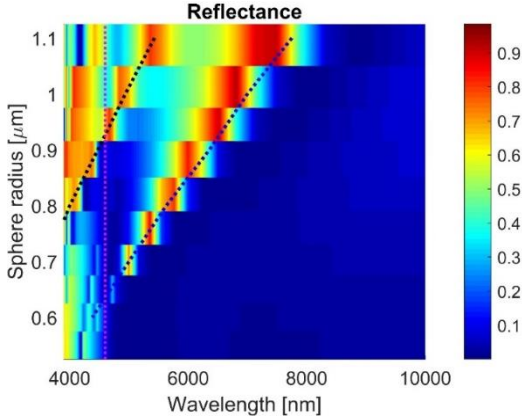

Figure 2. Reflectance in the array of HM bars and silicon spheres of varied sizes $R = 0.5 – 1.1$ µm. Dotted blue and black lines mark magnetic Si-MD and electric Si-ED dipole resonances of silicon sphere, respectively. The dotted magenta line corresponds to magnetic and electric quadrupole resonances (HM-MR/EQR) of the HM bar. HM bar dimensions are $a_x = 0.9$ µm, $a_y = 0.5$ µm, and $a_z = 2$ µm.

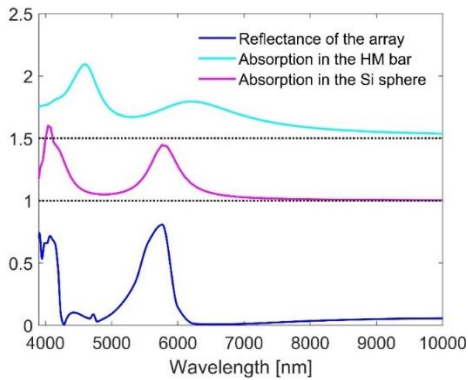

Figure 3. Absorption spectra for arrays of silicon spheres and HM bars as well as reflectance from the array of combined particles. The decrease of reflectance is observed in the spectral regions where of Si-MD overlaps with HM-EMR and Si-ED overlaps with HM-MR/EQR. Lines of absorption in silicon spheres and HM bars are shifted by 1 and 1.5 from the zero coordinate, respectively. HM bar dimensions are $a_x = 0.9$ µm, $a_y = 0.5$ µm, and $a_z = 2$ µm, and silicon sphere is $R = 0.82$ µm.